
\input phyzzx
\def\Bbb{\bf}

\def\zint{\oint\!{dz\over 2\pi i}\,}
\def\nicefrac#1#2{\hbox{${#1\over #2}$}}
\def\frac#1/#2{\leavevmode\kern.1em\raise.5ex
		\hbox{\the\scriptfont0
         	#1}\kern-.1em/\kern-.15em
		\lower.25ex\hbox{\the\scriptfont0 #2}}
\def\half{\nicefrac 1 2}

\pubnum={\vbox{\hbox{G\"oteborg-ITP-93-5}
		\hbox{hep-th/9303172}
			}}
\titlepage
\title{$N\!=\!8$ Superconformal Algebra and the Superstring}
\author{Lars Brink, Martin Cederwall\break\break
	 		{\it and}\break\break
		Christian R. Preitschopf}
\address{\break Institute for Theoretical Physics\break
	Chalmers University of Technology and University of G\"oteborg\break
	S-412 96 G\"oteborg, Sweden}
\abstract
The superstring in $D\!=\!3,4$ and 6 is invariant under
an $N\!=\!D\!-\!2$ superconformal algebra based on $\widehat{S^{D-3}}$.
There is a direct relationship between this (world-sheet) symmetry and
the super-Poincar\'e (target space) symmetry. We establish this
relationship using the light-cone gauge, show how the statement
generalizes to $D\!=\!10$ and examine the properties of the $N\!=\!8$
superconformal algebra and the possible implications of its existence.
\submit{{\sl Phys.Lett.}\bf B}
\endpage

\REF\Ven{G.~Veneziano\journal Nuovo Cim.&57A(68)190.}
\REF\Vir{M.A.~Virasoro\journal Phys.Rev.&D1(70)2933.}
\REF\Ram{P.~Ramond\journal Phys.Rev.&D3(71)2415.}
\REF\NS{A.~Neveu and J.H.~Schwarz\journal Nucl.Phys.&B31(71)86.}
\REF\GSO{F.~Gliozzi, J.~Scherk and D.~Olive\journal Nucl.Phys.&B122(77)253.}
\REF\GS{M.B.~Green and J.H.~Schwarz\journal Phys.Lett.&136B(84)367.}
\REF\kappasymm{W.~Siegel\journal Phys.Lett.&128B(83)397.}
\REF\secondclass{I.~Bengtsson and M.~Cederwall, G\"oteborg-ITP-84-21}
\REF\Metal{M.~Ademollo \etal\journal Phys.Lett.&62B(76)105,\
	{\it Nucl.Phys.} {\bf B111}(1976), 77,\nextline
	\indent{\it Nucl.Phys.}{\bf B114}(1976), 297\nextline
	and an enormous amount of later papers.}
\REF\LCstring{P.~Goddard, J.~Goldstone, C.~Rebbi and C.B.~Thorn,\nextline
	\indent{\it Nucl.Phys.} {\bf B56}(1973), 109,\nextline
	M.B.~Green, J.H.~Schwarz\journal Nucl.Phys.&
	B181(81)502.}
\REF\Sudbery{A.~Sudbery\journal J.Phys.&A17(84)939.}
\REF\Kugo{T.~Kugo and P.~Townsend\journal Nucl.Phys.&B221(83)357.}
\REF\sixtwist{I.~Bengtsson and M.~Cederwall\journal Nucl.Phys.&B302(88)81,
	\nextline M.~Cederwall\journal J.Math.Phys.&33(92)388.}
\REF\ManSud{C.A.~Manogue and A.~Sudbery\journal Phys.Rev.&D40(89)4073,
	\nextline C.A.~Manogue and J.~Schray, hep-th/9302044.}
\REF\quat{M.~Cederwall and C.R.~Preitschopf, G\"oteborg-ITP-92-40,
	hep-th/9209107,\nextline{\it Nucl.Phys.} {\bf B} in press.}
%
%
\REF\torsion{F.~G\"ursey and C.-H.~Tze, {\sl Phys.Lett.} {\bf 127B}(1983), 191;
	\nextline
	M.~Rooman, {\sl Nucl.Phys.} {\bf B238}(1984), 501.}
\REF\ESTPS{F.~Englert, A.~Sevrin, W.~Troost, A.~Van
	Proyen and Ph.~Spindel,\nextline\indent {\sl J.Math.Phys.}
	{\bf 29}(1988), 281.}
\REF\DTH{F.~Defever, W.~Troost and Z.~Hasiewisz\journal J.Math.Phys.
	&32(91)2285.}
\REF\eightcft{M.~Cederwall and C.R.~Preitschopf, {\it in preparation}.}
\REF\Berkovits{N.~Berkovits\journal Nucl.Phys.&B358(91)169.}
\REF\Osipov{E.P.~Osipov\journal Phys.Lett.&214B(88)371,\nextline
	\indent{\sl Lett.Math.Phys.} {\bf 18}(1989), 35.}

In order to understand the physics of a certain model, we need to find its
symmetries. The crucial step in the Veneziano model [\Ven]
was the discovery of the
Virasoro algebra [\Vir], the infinite-dimensional conformal symmetry.
It led to the
no-ghost theorem, and in the modern era it has been used \eg\ to construct
string field theory. It was also the extension of this symmetry to the
superconformal one by Ramond [\Ram] that led to the
Ramond-Neveu-Schwarz model [\Ram,\NS] we
use today. However, in this formalism global supersymmetry is somewhat hidden
and appears only after the GSO-projection [\GSO].
To remedy this fact, Green and
Schwarz proposed their formulation [\GS],
where space-time supersymmetry is manifest.
The superconformal symmetry, however, is unconventional [\kappasymm],
and leads to difficulties when one tries to quantize the theory covariantly
[\secondclass]. It is not directly related to the conventional extended
superconformal structures [\Metal].

In both these formulations, the target space symmetries are divorced from
the world-sheet ones. This is really not satisfactory. If there is a
``fundamental string theory'', we expect the two sets of symmetries to have
a common origin. There should be two equivalent principles
for formulating the theory.
On the one hand, starting with enough physical requirements on the
space-time theory, the world-sheet symmetries should follow, while on the
other hand the correct assumptions about the two-dimensional physics on the
world-sheet should give the right target space behaviour.

In this letter we will show that in the light-cone gauge there
is indeed such a correlation. The light-cone formulation can be obtained
by using either principle.
The resulting theory is described by the transverse coordinates and is
explicitely unitary. The difficulties arise in the (super-)Poincar\'e
algebra which is non-linearly implemented [\LCstring].
In the quantum case anomalies
occur, unless the critical dimension is chosen.

The Green-Schwarz string [\GS] in the light-cone gauge is
described by an action
(in a heterotic form)
$$S={\nicefrac 1\pi}\!\int\! dzd\overline z\,
 (\partial\varphi^I\overline\partial
\varphi^I+S^a\overline\partial S^a)\eqn\lightconeaction$$
The index $I\!=\!0,\ldots ,D\!-\!3$ is vectorial,
while $a\!=\!0,\ldots ,D\!-\!3$ is spinorial.
The action \lightconeaction\ is classically super-Poincar\'e invariant for
$D\!=\!3,4,6$ and 10 [\LCstring].
Quantum mechanically matter has to be added for $D\!=\!3,4$ and 6 in order to
avoid anomalies.
Already at this point we would like to remark that the potential occurrence
of an anomaly in the super-Poincar\'e algebra points towards a close
connection between some of its generators and the world-sheet symmetry
that carries the anomaly in a covariant formulation. The most important issue
of this paper is to specify this connection.

Before going into details on the super-Poincar\'e and superconformal
algebras, we would like to introduce some division algebra formalism,
by now well known to be related to the space-times with $D\!=\!3,4,6$ and 10
and to the supersymmetric structures appearing in these dimensionalities
[\Sudbery
	-\quat].
We denote by ${\Bbb K}_\nu$ the division algebra of dimension $\nu$:
${\Bbb K}_1\!=\!{\Bbb R}$, the reals, ${\Bbb K}_2\!=\!{\Bbb C}$, the complex
numbers, ${\Bbb K}_4\!=\!{\Bbb H}$, the quaternions, and ${\Bbb K}_8\!=\!
{\Bbb O}$, the
octonions.
In the following, $\nu$ and $D\!-\!2$ are exchangeable.
Conjugation of an element $x\!\in\!{\Bbb K}_\nu$ is denoted $x^*$, not to
confuse
with the complex structure of the world-sheet.
Division algebra multiplication encodes the Clifford algebra of transverse
space-time. So for example, is the equation $c^*\!=\!vs$ equivalent to
$c_{\dot a}\!=\!v^I\gamma^I_{\dot aa}s_a$, where $v\!\in\! 8_v$,
$s\!\in\! 8_s$ and
$c\!\in\! 8_c$ of $SO(8)$, and analogously for lower dimensionalities.
We also use the notation $[x]=\half(x\!+\!x^*)$ and $\{x\}=\half(x\!-\!x^*)$.
Structure constants and associator coefficients are defined by
$[e_i,e_j]=2\sigma_{ijk}e_k$, $[e_i,e_j,e_k]=2\rho_{ijkl}e_l$, where $\{e_i,
i=1,\ldots,7\}$ are the imaginary units.

The action \lightconeaction\ can trivially be rewritten in this notation as
$$S={\nicefrac 1\pi}\!\int\! dzd\overline z\,[\,\partial\varphi^*
	\overline\partial\varphi+
	S^*\overline\partial S\,]\eqn\divaction$$

In $D\!=\!3,4$ and 6, the light-cone superstring action is invariant under an
$N\!=\!D\!-\!2$ extended superconformal algebra [\quat]. We only consider
the part of the generators containing holomorphic fields, with
$$\eqalign{\varphi^I(z)\varphi^J(\zeta)&\sim\delta^{IJ}\ln(z-\zeta)\cr
S^a(z)S^b(\zeta)&\sim{\delta^{ab}\over z-\zeta}\cr}\eqn\correlators$$
The generators of the algebras are [\quat]
$$\eqalign{\cal J&=\half S^*\!S	\cr
	\cal G&=\partial\varphi S	\cr
	\cal L&=\half[\partial\varphi^*\partial\varphi-S^*\partial S]\cr}
				\eqn\SCA$$
with the operator product expansions
$$\eqalign{{\cal J}_\alpha(z){\cal J}_{\alpha'}(\zeta)\;&\sim\;
		{\frac c/3\over(z-\zeta)^2}[\alpha\alpha']+
		{2\over z-\zeta}{\cal J}_{\alpha\alpha'}	\cr
	{\cal J}_\alpha(z){\cal G}_\Omega(\zeta)\;&\sim\;
		-{1\over z-\zeta}{\cal G}_{\Omega\alpha}	\cr
	{\cal G}_\Omega(z){\cal G}_{\Omega'}(\zeta)\;&\sim\;
		{\frac 2c/3\over(z-\zeta)^3}[\Omega^*\!\Omega']+
		{2\over(z-\zeta)^2}{\cal J}_{{\Omega}^*\Omega'}(\zeta)+
		{1\over z-\zeta}(\partial{\cal J}_{{\Omega}^*\Omega'}+
		2\,[\Omega^*\!\Omega']{\cal L})			\cr
	{\cal L}(z){\cal J}(\zeta)\;&\sim\;
		{1\over (z-\zeta)^2}{\cal J}(\zeta)+
		{1\over z-\zeta}\partial{\cal J}		\cr
	{\cal L}(z){\cal G}(\zeta)\;&\sim\;
		{\frac 3/2\over (z-\zeta)^2}{\cal G}(\zeta)+
		{1\over z-\zeta}\partial{\cal G}		\cr
	{\cal L}(z){\cal L}(\zeta)\;&\sim\;
		{\frac c/2\over(z-\zeta)^4}+
		{2\over (z-\zeta)^2}{\cal L}(\zeta)+
		{1\over z-\zeta}\partial{\cal L}		\cr}
			\eqn\superOPE$$
where fields in $\nu$- or $(\nu\! -\!1)$-dimensional representations are
given indices by contractions $X_a=[a^*X]$, and where the anomaly $c$ with
these minimal field contents take the value $3\nu/2$.
The Kac-Moody part of this
algebra is $\widehat{S^{\nu -1}}$. A similar statement applies to $D\!=\!10$,
as we will soon describe. By examining the gauge-fixing procedure
of the Green-Schwarz superstring~[\GS] to the light-cone
gauge, and in particular the non-linear realization of the
super-Poincar\'e algebra, we will give an interpretation of the
world-sheet superconformal algebra in terms of space-time supersymmetry.

Consider the constraints derived from the Green-Schwarz action for the
left-moving variables of a heterotic string (in $SL(2;{\Bbb K}_\nu)$-notation):
$$\eqalign{D_\alpha&\equiv\pi_{\dot\alpha}-{1\over\sqrt 2}
		\Pi_{\dot\alpha\alpha}\theta^\alpha+
		\nicefrac{1}{4}(\theta\partial\theta^\dagger-
		\partial\theta\theta^\dagger)_
		{\dot\alpha\alpha}\theta^\alpha\approx 0\cr
	L&\equiv{\nicefrac 1 4}\Pi_{\dot\alpha\alpha}\Pi^{\alpha\dot\alpha}
				\approx 0\cr}\eqn\GSconstr$$
where $\Pi=\partial\varphi+{1\over\sqrt 2}(\theta
\partial\theta^\dagger-\partial\theta\theta^\dagger)$ and $\pi$ is the
conjugate momentum of $\theta$. The special property of
the spinorial constraint is that it contains an equal number of
first and second class constraints [\kappasymm,\secondclass]. When
a light-cone gauge is chosen, the separation comes about
naturally,
splitting the $SO(1,D-1)$ spinor into two spinors of the transverse
group. The light-cone gauge amounts to choosing
$$\eqalign{\varphi^+(z)&=\alpha^+\ln z		\cr
	\theta^1&=0				\cr}\eqn\LCgauge$$
The remaining part of the spinorial constraint reads
$\pi^2+{\alpha^+\over z}\theta^2\approx 0$ and is obviously second class.
When we eliminate it and define $S=\sqrt{2\alpha^+\over z}\theta^2$,
the spinor correlator in eq.\correlators\ is recovered.
Then, one can solve for $\varphi^-$ and $\pi^1$ through eq.\GSconstr\
to obtain
$$\eqalign{\partial\varphi^-&={z\over 2\alpha^+}[\partial\varphi^*
			\partial\varphi-S^*\partial S]			\cr
	\pi^1&={1\over 2}\sqrt{z\over\alpha^+}\partial\varphi S	\cr}
					\eqn\solvvar$$

When we now go back to the superconformal generators of eq. \SCA, we
notice that the variables eliminated by the gauge choices are
exactly the superconformal generators $\cal L$ and $\cal G$.
{}From the light-cone variables, we can construct the now non-linearly
realized super-Poincar\'e generators. The crucial part, \ie\
the part where anomalies may appear, contains $P^-, J^{+-}, J^-$ and
$Q^-$, the generators that take us
out of the quantization surface. The complete set of generators is
[\LCstring]
$$\eqalign{P^+\;&=\;\alpha^+						\cr
	P^-\;&=\;{1\over 2\alpha^+}\zint z\,[\,\partial\varphi^*
		\partial\varphi-S^*\partial S\,]={{\cal L}_0\over\alpha^+}  \cr
	P\;&=\;p							\cr
	J^{+-}\;&=\;x^+{1\over 2\alpha^+}\zint z\,[\,\partial\varphi^*
		\partial\varphi-S^*\partial S\,]+\alpha^+\partder{}{\alpha^+}=
		{x^+\over\alpha^+}({\cal L}_0-\half)
		+\alpha^+\partder{}{\alpha^+}   			\cr
	J^+\;&=\;x^+p-\alpha^+x						\cr
	J^-\;&=\;-p\partder{}{\alpha^+}-{1\over 2\alpha^+}
		\zint z\,\Bigl(\tilde\varphi\,
		[\,\partial\varphi^*\partial\varphi-S^*\partial S
			-{1\over z^2}]
		-\half(\partial\varphi S)S^*\Bigr)=			\cr
		&=\;-p\partder{}{\alpha^+}-{1\over\alpha^+}\zint z\Bigl(
		\tilde\varphi({\cal L}-{1\over 2z^2})
		-\nicefrac{1}{4}{\cal G}S^*\Bigr)			\cr
	J^{IJ}\;&=\;2x^{[I}p^{J]}+\zint\Bigl(\tilde\varphi^I\partial
		\tilde\varphi^J+\nicefrac{1}{4}[S^*e^*_I(e_JS)]\,\Bigr)	\cr
	Q^+\;&=\;2^{1/4}\sqrt{\alpha^+}\zint z^{-1/2}S			\cr
	Q^-\;&=\;{1\over 2^{\frac 1/4}\sqrt{\alpha^+}}\zint z^{1/2}
		\partial\varphi S={1\over 2^{\frac 1/4}\sqrt{\alpha^+}}
		{\cal G}_0		\cr}\eqn\superP$$
(transverse indices are again suppressed, so that $J^-$ contains
the components $J^{-I}$ \etc). For convenience, we have separated out
the logarithmic mode of $\varphi^\mu$ according to $\tilde\varphi^\mu(z)
=\varphi^\mu(z)-\ln z\oint\!{d\zeta\over 2\pi i}\,\partial\varphi^\mu(\zeta)$,
and defined $x^\mu=\oint\!{dz\over 2\pi iz}\,\tilde\varphi^\mu(z)$.
The important point is that
knowledge of the super-Poincar\'e generators provides us with information
about the superconformal generators, and vice versa.
Explicit calculation of the anomaly in $[J^{-I},J^{-J}]$ of course gives
the result $\nu\!=\!8$.

Compactification to $D\!=\!6$ yields the following changes:
$$\eqalign{
	P^-\;&={{\cal L}_0 + {\cal L}_0^{int} \over\alpha^+}  \cr
	J^{+-}\;&={x^+\over\alpha^+}({\cal L}_0 + {\cal L}_0^{int} -\half)
		+\alpha^+\partder{}{\alpha^+}   			\cr
	Q^-\;&=\;{1\over 2^{\frac 1/4}\sqrt{\alpha^+}}
		({\cal G}_0 + {\cal G}_0^{int} )  \cr
	J^-\;&=\;-p\partder{}{\alpha^+}-{1\over\alpha^+}\zint z\Bigl(
		\tilde\varphi({\cal L}+ {\cal L}^{int}-{1\over 2z^2})
		-\nicefrac{1}{4}({\cal G} + 2 {\cal G}^{int} )S^*+\cr
&\phantom{=\;-p\partder{}{\alpha^+}-{1\over\alpha^+}\zint z\Bigl(
		\tilde\varphi({\cal L}+ {\cal L}_0^{int}-{1\over 2z^2})\;\,}
		+ \half {\cal J}^{int}\partial\varphi
		 \Bigr) \cr
	J^{IJ}\;&=\;2x^{[I}p^{J]}+\zint\Bigl(\tilde\varphi^I\partial
		\tilde\varphi^J+\nicefrac{1}{4}[S^*e^*_I(e_JS)]
		+\half[e^*_I e_J {\cal J}^{int}]\,\Bigr)	\ \ \cr}
\eqn\compsix
$$
where ${\cal J}^{int}$, ${\cal G}^{int}$ and ${\cal L}^{int}$
is an $N\!=\!4$ superconformal algebra for the internal degrees of freedom,
and we are working with quaternions instead of octonions.
An anomaly-free theory arises if $c^{int}=6$ and if the nullmode condition
$$
\zint z \Bigl( \ {\cal J}^{int}_I  {\cal J}^{int}_K - \coeff{1}{3}
\delta_{IK}  {\cal J}^{int}_L  {\cal J}^{int}_L \Bigr)\ = 0\
\eqn\nullmode
$$
is satisfied. We note that while ${\cal J}= \half S^*\!S$ contains the
antiselfdual combination of spinors, we find in $J^{IJ}$ the selfdual
combination. Hence the internal algebra has the structure corresponding
to an antiselfdual multiplet, while the transverse spacetime algebra
corresponds to a selfdual multiplet. These two $N\!=\!4$ algebras are
independent, and when one goes about constructing sigma-models,
one will have to consider two independent hyperk\"ahler structures, one in
the internal sector and one in the noncompact sector.

If we compactify down to $D\!=\!4$, we obtain
the same operators as in \compsix, except for $J^-$, which now reads
$$\eqalign{
J^-\ =\ -p\partder{}{\alpha^+}-{1\over\alpha^+}\zint z\Big(
	& \tilde\varphi({\cal L}+ {\cal L}^{int}-{1\over 2z^2})
		-\nicefrac{1}{4}({\cal G} + 2 {\cal G}^{int} )S^*
	\cr &\
	- \half {\cal J}^{int}\partial\varphi
		+ \half {\cal A}^* \Big) \cr
}
\eqn\compfour
$$
Now the internal algebra is more complicated. It contains
the $N\!=\!2$, $c\!=\!9$ superconformal algebra
${\cal J}^{int}$, ${\cal G}^{int}$ and ${\cal L}^{int}$
as well as a complex chiral multiplet $({\cal A}$, ${\cal R})$
of conformal weights $(2,\frac3/2)$ with the following operator products:
$$
\eqalign{
{\cal G}^{int}_a(z) {\cal G}^{int}(\zeta) \ &\sim \
	{6 e_a\over (z-\zeta)^3} + {2 e_a\over z-\zeta} {\cal L}^{int}
	- {e_a\over (z-\zeta)^2} \big({\cal J}^{int}(z) +
	{\cal J}^{int}(\zeta)\big)
\cr
{\cal J}^{int}(z) {\cal G}^{int}(\zeta) \ &\sim \
	-{i\over z-\zeta} {\cal G}^{int}
\cr
{\cal J}^{int}(z) {\cal J}^{int}(\zeta) \ &\sim \
	{1\over (z-\zeta)^2}
\cr
 & \cr
{\cal G}^{int}_a(z) {\cal A}(\zeta) \ &\sim \
	{3 e_a^*\over (z-\zeta)^2} {\cal R}(\zeta) + { e_a^*\over z-\zeta}
	\partial{\cal R}
\cr
{\cal G}^{int}_a(z) {\cal R}(\zeta) \ &\sim \
	{ e_a \over z-\zeta} {\cal A}
\cr
{\cal J}^{int}(z) {\cal A}(\zeta) \ &\sim \
	-{i\over z-\zeta} {\cal A}
\cr
{\cal J}^{int}(z) {\cal R}(\zeta) \ &\sim \
	-{2i\over z-\zeta} {\cal R}
\cr
 & \cr
{\cal R}_a(z) {\cal R}(\zeta) \ &\sim \
	-{4 e_a\over (z-\zeta)^3}
	+ {2 e_a \over (z-\zeta)^2}
		\big({\cal J}^{int}(z) + {\cal J}^{int}(\zeta)\big)
	\cr & \phantom{mmm}
		- {32 e_a\over z-\zeta} :({\cal J}^{int})^2:
\cr
{\cal A}_j(z) {\cal R}(\zeta) \ &\sim \
 	  {2 e_j \over (z-\zeta)^2}{\cal G}^{int}(\zeta)
	+ {2 e_j\over z-\zeta} : {\cal G}^{int}{\cal J}^{int}:
\cr
{\cal A}_j(z) {\cal A}(\zeta) \ &\sim \
	-{ 12 e_j\over (z-\zeta)^4}
	+{ 4 e_j^*\over (z-\zeta)^3}  \
		\big({\cal J}^{int}(z) + {\cal J}^{int}(\zeta)\big)
	\cr &\phantom{mmm}
	 -{ e_j\over (z-\zeta)^2}
		\big( (:({\cal J}^{int})^2(z): + {\cal L}^{int}(z)) +
		(z\leftrightarrow \zeta) \big)
	\cr &\phantom{mmm}
	+ { e_j^*\over z-\zeta} \big( 3\partial^2 {\cal J}^{int} -
		4: {\cal L}^{int}  {\cal J}^{int}:
		- :({\cal G}^{int})^2: \big)
\cr
}
\eqn\fouralg
$$
Expressions like $:({\cal G}^{int})^2:$ in the above expressions
are normal ordered with respect to the modes of the currents. This
prescription differs from the normal ordering with respect to the
modes of, say, free component fields.
The operator algebra in \fouralg\ replaces the nullmode
condition \nullmode. We do not know whether it is peculiar to some
compactifications to $D\!=\!4$ or a more general property of
the internal sector of $D\!=\!4$ superstrings.
We have here an algebraic
structure on the internal space without explicit appearance of coordinates.
We have not checked, but it may well be possible to do so, whether this
algebra, or some algebra of this type {\it has} to appear for $J^-$
to be non-anomalous. If that is the case, one will have an instrument
for treating the internal manifold in an abstract algebraic manner,
that might be useful for extracting the physical consequences of
specific choices for internal manifolds, and possibly for demonstrating
``equivalence'' between different manifolds with respect to their
properties concerning string propagation.

What we still have not shown is that the interpretation
of the super-Poincar\'e generators in terms of superconformal generators
is valid also for the case $D\!=\!10$. That will be the subject of the
rest of this letter.

Let us now turn to the generalization to $N\!=\!8$.
We will give an intuitive step by step construction leading to the final
form of the $N\!=\!8$ superconformal generators and their algebra.
Since on the light cone
the spacetime supersymmetry algebra for $D\!=\!10$ has the same structure as
for $D\!=\!6$ without a compact sector, one feels compelled to simply
replace quaternions with octonions. The operator product of the
imaginary currents $J=\coeff{1}{2}S^*\!S$ is then given by
$$
J^i(z)J^j(\zeta)\ =\ -{4\over(z-\zeta)^2}\  \delta^{ij}\ +\ {2\over
z-\zeta}\ \Big( \sigma_{ijk}\ J^k +  \rho_{ij\alpha\beta}
S^\alpha S^\beta   \Big) \eqn\JJ
$$
Hence this current algebra does not close, and we may attribute this
fact to the nonassociativity of the octonions, or equivalently to the
fact that $S^7$ is not a group manifold. Using octonions, the 7-sphere
is economically described by $S^7=\{X \in {\Bbb O}\ | \ |X|=1\}$,
with tangent vectors $Xe_i$ and normal $X$ [\torsion]. This defines a
connection without curvature and torsion $ T_{ijk}(X) =
[(Xe_i)^*(Xe_j) e_k] $. Note that
$$ T_{ijk}(X)|_{X=1} =
\sigma_{ijk} \qquad
\nabla_i T_{jkl}(X)|_{X=1} = 2 \rho_{ijkl}\eqn\structure
$$
Hence we may move from the north pole $X\!=\!1$ and form $J$ by multiplication
in a basis corresponding to another point on $S^7$,
$J=\coeff{1}{2} (XS)^*(XS)$, to obtain
$$
J^i(z)J^j(\zeta)\ =\ -{4\over(z-\zeta)^2}\  \delta^{ij}\ +\ {2\over
z-\zeta}\ \Big( T^{ijk}(X)\ J^k +  \nabla^i J^j \Big) \eqn\JJJ
$$
The rest of the algebra has a similar structure:
for $G=(X\varphi)^*(XS) = \sigma^I_{ab}(X) S_b \varphi^I $,
with $\sigma^I_{ab}(X) = [e^I(e_aX^*)(Xe_b^*)]$,
we obtain an algebra that closes modulo infinitesimal shifts on $S^7$, \ie\
besides the ``expected'' terms there are terms containing $\nabla^i$.
By considering finite
transformations generated by $J$, we see that we transform $Xe_a
\rightarrow Y(Xe_a)$ for another unit octonion $Y$, i.e. we obtain a
rotated basis of tangent vectors at $YX \in S^7$. Clearly we can also
generate the basis $(X^* Y^*) (Y(Xe_i))$ at the northpole $Z=1$.  This
basis is rotated with respect to the $Xe_i$. We conclude that the
shifts operate not on the 7-sphere, but on an SO(7)-bundle over $S^7$,
i.e. on SO(8). The operator product of, say, $J_X= \coeff{1}{2}
(XS)^*(XS)$ and $J_{YX}=\coeff{1}{2}(Y(XS))^*(Y(XS))$, does
not fit into the simple scheme displayed above. But then, we would
expect it to contain terms with an infinite number of
$S^7$-derivatives.  We will call the structure we found a current
algebra that is ``local on $S^7$''.

Up to this point we have treated $X$ as a number.
For the algebra to close, we need a mechanism that takes care of the
infinitesimal shifts on $S^7$. This is accomplished by letting the
$S^7$ coordinate $X$ be an operator, and adding an $S^7$ translation
generator to $J$. More precisely,
we introduce a pair of octonionic bosons $(\lambda, \omega)$ with
conformal weights $(\frac1/2,\frac1/2)$ and their superpartners $(\theta,\pi)$
of weights $(0,1)$, set $X=\lambda/ |\lambda|$ and define
$$
\eqalign{
&{\cal J} = \{\omega^*\lambda\}+\coeff{1}{2} (XS)^* (XS)
\cr
&{\cal G} = \pi^* \lambda - \partial\theta^*\omega + (X\partial\varphi)^*
(XS) + \coeff{1}{2} (XS)^* (\Lambda S) -
\coeff{1}{2} (\Lambda S)^* (XS) \cr
&{\cal L} = \coeff{1}{2} [\partial \lambda^*  \omega-\lambda^* \partial \omega]
- [\pi^*\partial\theta]
+\coeff{1}{2} [\partial\varphi^* \partial\varphi]
- \coeff{1}{2}[S^*\partial S]
\cr}\eqn\eightgener
$$
where $\Lambda = |\lambda|^{-1} (\partial\theta - X[X^*\partial\theta])$
is the tangential part of $\partial\theta$.
The algebra of these operators is soft [\ESTPS], i.e. it closes with
field dependent structure ``constants'' and anomaly terms.
The classical algebra is
$$\eqalign{{\cal J}_\alpha(z){\cal J}_{\alpha'}(\zeta)&\;\sim\;
		{2\over z-\zeta}{\cal J}_{(\alpha X^*)(X\alpha')}	\cr
	{\cal J}_\alpha(z){\cal G}_\Omega(\zeta)&\;\sim\;
		-{1\over z-\zeta}
		\Bigl({\cal G}_{(\Omega X^*)(X\alpha)}
		+{\cal J}_{\lambda^{-1}\bigl((\partial\theta\Omega)\alpha
		-\partial\theta((\Omega X^*)(X\alpha))\bigr)}\Bigr)\cr
	{\cal G}_\Omega(z){\cal G}_{\Omega'}(\zeta)&\;\sim\;
		{2\over(z-\zeta)^2}{\cal J}_{(\Omega^*X^*)(X\Omega)}
			+{2\over z-\zeta}\Bigl(\half\partial({\cal J}_
			{(\Omega^*X^*)(X\Omega')})
			+[\Omega^*\Omega']{\cal L}\Bigr)		\cr}
							\eqn\eightcorr$$
(the quantum algebra involves some subtleties that will be addressed in
a forthcoming paper [\eightcft]).
We note that only $\lambda$ and $\theta$ enter into the field
dependence, so that the structure functions (anti)commute. They have
a natural interpretation in terms of the torsion tensor and its
superpartner on $S^7$. The reduction to the $N\!<\!8$ algebras of
eq.\superOPE\ is obvious: just remove all associator terms. If one
replaces the term $\coeff{1}{2} (XS)^* (XS)$ in $\cal J$ by
$\{(X\omega')^* (X\lambda')\}$, where $(\lambda',\omega')$ is a conjugate
pair of bosons of weights (\frac1/2,\frac1/2), and makes
the corresponding replacements in $\cal G$ and $\cal L$, one finds the soft
algebra Berkovits describes in the context of the twistor formulation
of the superstring [\Berkovits].
Thus we have found a natural
generalization of the N=4 free field constructions.

We want to emphasize that we are working with explicit generators, and
therefore automatically have the Jacobi identities fulfilled. If we on
the right hand side of eq. \eightcorr\ set $X\!=\!1$, $\partial\theta\!=\!0$,
we get a non-associative
algebra like the ones in [\ESTPS,\DTH,\Osipov].
The present formulation is stronger.
A non-trivial feature is that, unlike what could be expected from a naive
consideration of the properties of the octonions,
the Kac-Moody part $\widehat{S^7}$ actually
commutes with the $SO(8)$ of space rotations, and that our seven-sphere
is therefore not the quotient of this group with an $SO(7)$ subgroup, but
an additional symmetry. This and other issues concerning the $N\!=\!8$
algebra are to be developed in detail in a forthcoming publication [\eightcft].

The $N\!=\!8$ generators of eq. \eightgener\ as they stand are not the
ones that enter in the super-Poincar\'e generators \superP.
First the ``parameter fields'' $(\lambda,\omega)$ and $(\theta,\pi)$ must be
removed --- they are not physical fields. With our present understanding of
the role of these fields we cannot make any certain statements about
their physical interpretation in a covariant theory. We do not for example
know what the constraints are that eliminate the parameter fields.
A possible interpretation is that they are a remnant of a set of
super-twistor variables from a combined space-time/twistorial formulation.
For the moment we will take a very pragmatic point
of view and note that in order to reduce the field content to that of
the light-cone superstring, we need some quantum mechanically consistent
set of constraints (note that the superconformal generators cannot
be set to zero with a quantum-mechanically nilpotent BRST charge).
We may state $\omega\approx 0$ and $\pi\approx 0$,
allowing for the gauge choices $X\!=\!1,\ \theta\!=\!0$,
which of course takes us back
to the situation in eqs.~\JJ\ and \JJJ.
The closure of the algebra is obstructed
by the gauge choices. However, the role of the generators of the
superconformal algebra in the super-Poincar\'e algebra is identical
to that in the lower dimensionalities.

Finally, one may speculate in the ultimate role of the superconformal
algebra in some kind of ``covariant'' formulation. We have a strong
belief that the $N\!=\!D\!-\!2$ superconformal algebras have a fundamental
significance, yet their generators enter very asymmetrically \eg\ in
the super-Poincar\'e generators, where ${\cal J}$ is not seen at all.
It is tempting to think that the relation between space-time and
worldsheet symmetries established in this paper gives a glimpse
of the structure of a bigger symmetry.

\ack
We thank Nathan Berkovits for discussions on the
$N\!=\!8$ algebra, and Viktor Ogievetsky for pointing out some
references.

\refout
\end